\documentclass[12pt,a4paper]{article}

\usepackage{epsfig}

\newcommand{\beq}[1]{\begin{equation}\label{#1}} 
\newcommand{\eeq}{\end{equation}} 
\newcommand{\beqar}[1]{\begin{eqnarray}\label{#1}} 
\newcommand{\eeqar}{\end{eqnarray}}

\author{P.~H\"agler$^1$, A.~Sch\"afer$^2$}               

\title{Evolution equations for higher moments of
 angular momentum distributions} 

\date{} 

\begin{document}

\maketitle 
 
\begin{center} 
$^1$Philipp H\"agler, c/o Prof. Andreas Sch\"afer\\
Institut f\"ur Theoretische Physik\\
Universit\"at Regensburg\\
D-93040 Regensburg, Germany\\
E-Mail: haegler@th.physik.uni-frankfurt.de\\ 
$^2$Prof. Andreas Sch\"afer\\
Institut f\"ur Theoretische Physik\\
Universit\"at Regensburg,    
D-93040 Regensburg, Germany\\ 
E-Mail: andreas1.schaefer@physik.uni-regensburg.de\\
Telephone: (49) - 941 - 943 - 2007\\
FAX:       (49) - 941 - 943 - 3887\\
\end{center} 
\vskip 0.3cm
PCAS: 12.38.Bx\\
\begin{tabular}{ll}
Keywords:&perturbative QCD, orbital angular momentum,\\
&evolution equation, nucleon spin\\
\end{tabular}
\vskip 0.3cm
\noindent{\bf Abstract:}   
\vskip 0.3 cm 
Based on a sumrule for the nucleon spin we
expand quark and gluon orbital angular momentum operators
and derive an evolution matrix
for higher moments of the corresponding distributions.
In combination with the spin-dependent DGLAP-matrix 
we find a complete set of spin and orbital angular momentum
evolution equations.

 \eject 
\newpage 
\section {Introduction} 
Since several years much effort is invested into the determination of 
the internal nucleon spin
structure by means of  polarized lepton-nucleon deep-inelastic-scattering.
One of the focal questions in this respect is the determination of the
gluon spin fraction $\Delta G(x,Q^2)$, which could be as large as 2,
though its actual value is only poorly known. Let us refere here just
to some recent analyses 
\cite{NLO1,NLO2,NLO3,NLO4,NLO5}
instead of reviewing the slightly involved
history of this discussion.
As the total angular momentum of the nucleon is one half and
the quark spin is known experimentally to be small, a large
gluon spin implies substantial angular momentum. 
It is a natural hope that angular distributions of final state
hadrons can be linked to the angular momenta of quarks and
gluons inside of the struck nucleon. Up to know it was, however, not
possible to establish this connection within a self-consistent 
QCD scheme. This would require e.g. the correct treatment of
transverse momentum, including  all spin-momentum-correlations, 
 during the fragmentation process. \\ 
Here we want to disregard such problems and focus instead on 
the combined evolution equations for quark and gluon spin
and angular momentum distributions. This evolution was studied for
the first moment quite extensively by e.g. Xiangdong Ji \cite{JiI}.
We extend his analyses to higher moments. The relevance of such
investigations for the possibilities to find experimental observables 
linked to the partonic angular momentum is probably rather obvious.
The $x$ and $Q^2$ dependence of the angular momentum distributions 
will be a crucial ingredient to decide 
whether it is possible to find for existing or planned
experiments an $x$-$Q^2$-window for which simultaneously $Q^2$ is
large 
enough for perturbative QCD to apply, $\nu$ is large enough to assume 
leading quark fragmentation, and the spin-asymmetries are so large,
that an experimental determination seems possible.
\section{Preliminaries} 
Let us begin with a presentation of a partition formula  
for the spin of a nucleon in an helicity-eigenstate in 
light-cone coordinates (LCC).  
The angular momentum density is defined through
 \begin{equation}
 M^{\mu\nu\lambda}(x) = x^\nu T^{\mu\lambda}(x
 )-x^\lambda T^{\mu\nu}(x),
\label{Mmunulambda}
\end{equation} 
which is a gauge invariant rank-$3$ tensor.
Let now $ | P,S \rangle $ be a
nucleon-state with the normalization
\begin{equation}
 \langle P',S' | P,S \rangle = 
 2P^+(2\pi)^3\delta^3(\vec{P}-\vec{P'})
 \delta_{SS'}.
\label{norm}
\end{equation}
In the following we consider the quantity 
\begin{equation}
 \widetilde{M}^{\mu\nu\lambda}(P,P',S) = 
 \int d^3x\, \langle P',S |M^{\mu\nu\lambda}(x)| P,S \rangle 
\label{Mschlange}
\end{equation}
and treat it as a distribution in the variable  
$\vec{P'}$ (in LCC, the integral becomes $\int
dx^-dx^1dx^2$). To avoid derivatives of
delta distributions or delta distributions  
at the point 0, a bilinear functional 
\begin{equation}
\bigg(f(P,P'),\Phi (P')\bigg) = \int d^3 \vec{P}' f(P,P')
 \Phi (P')
\label{skalarprod}
\end{equation}
is defined with a suitable test function $\Phi(P')
$ for an arbitrary distribution $f(P,P')$. A calculation  
quite similar to the one in \cite{JaffeI} leads to  
\begin{equation}
 \frac {\bigg( \widetilde{M}^{+12}(P,P',S), \Phi (P')\bigg)} 
       {2P^+ (2 \pi)^3 \Phi (P)}
       =
 \frac {1} {2},
\label{summenregel}
\end{equation}
where $P^\mu=(P^+,0,0,P^-)$, $S^\mu=(P^+,0,0,-P^-)$, and  
\begin{equation}
 \bigg(\langle P',S | P,S \rangle,\Phi (P')\bigg) =
 2P^+ (2 \pi)^3 \Phi (P)
\end{equation} 
plays the role of the norm (\ref{norm}) . 
In the case of QCD, the above component of the  
angular momentum density tensor is given by
(in the following we use the light cone gauge 
$A^+=0$) 
\begin{eqnarray}
 M^{+12}(x)&=&
 i\overline{\Psi}\gamma^+(x^1
 \partial^2-x^2\partial^1)\Psi \nonumber\\  
 && + \partial^+A^j(x^1\partial^2-x^2\partial^1
 )A^j \nonumber\\   
 && + \frac {i} {4} \overline{\Psi}\gamma^+[\gamma^1,
 \gamma^2]\Psi \nonumber\\   
 && + A^1\partial^+A^2 - A^2\partial^+A^1,
\label{MQCD}
\end{eqnarray} 
where j is summed over 1 and 2; and  $\partial^+ = \frac {\partial} 
{\partial x_+} = \frac {\partial}{\partial x^-} = \partial_-$. 
Using the definitions (see \cite{JiI,JaffeI}) 
\begin{eqnarray}
 \Delta\Sigma &=& \langle P',S|\int d^3x\,
  \frac{i}{2} \overline{\Psi}\gamma^+[\gamma^1,
  \gamma^2]\Psi |P,S\rangle 
 \label{deltasigma} \\ 
  \Delta g &=& \langle P',S|\int d^3x\, (A^1
  \partial^+A^2 - A^2\partial^+A^1)|P,S\rangle 
 \label{deltag} \\
 L_q &=& \langle P',S|\int d^3x\, i \overline{\Psi}
  \gamma^+ (x^1\partial^2-x^2\partial^1)\Psi |P,S
  \rangle 
 \label{Lq} \\
  L_g &=& \langle P',S| \int d^3x\, \partial^+
 A^j (x^1\partial^2-x^2\partial^1) A^j
 |P,S \rangle
 \label{Lg}
\end{eqnarray} 
and (\ref{summenregel}), one gets 
\begin{equation}
 \frac {\bigg(\Delta\Sigma/2 + \Delta g + L_q + L_g,
  \Phi(P')\bigg)}{2P^+(2\pi)^3\Phi(P)} = \frac{1}{2}.
\label{summezerlegt}
\end{equation}
This is the nucleon spin sumrule.
Because the 3-space integral over $M^{+12}(x)$ is equal to the
third component of the generator of rotations, (\ref{summezerlegt}) can
be seen as a decomposition of the nucleon spin into quark and gluon spin
and orbital angular momentum contributions
\cite{JiIII}. $\Delta g$ turns out to be the first moment
of the difference of the gluon
distributions with longitudinal spin projection $\pm 1$.
\begin{equation}
 \Delta g(t)=\int^1_0 dx \Big[g_\uparrow(x,t) -
 g_\downarrow(x,t)\Big].
\label{deltag2}
\end{equation}
Due to the general scale dependence of the theory,
the quark and gluon distributions are functions
of $t=\ln(Q^2/\Lambda_{QCD}^2)$ ($Q^2$ being the typical scale
set by  the
external momenta).
The t evolution of the n-th moments of the spin
dependent parton distributions, $\Delta
g^n(t)$ and $\Delta\Sigma^n(t)$, is given
by the well known
DGLAP-equation (\cite{AP},\cite{A}). The evolution
matrix is the
anomalous dimension matrix of the corresponding
operators, e.g. for
the quark spin operator (in normal coordinates)
\begin{eqnarray}
 \widehat S_q&=&\int d^3x\, \frac{i}{2} \overline{
  \Psi}\gamma^0[\gamma^1,
  \gamma^2]\Psi = \int d^3x\,\overline\Psi\gamma^3\gamma_5
  \Psi \nonumber\\
 \longrightarrow \widehat{S}^{\mu_1\ldots
  \mu_{n-1}}_q&=&\int d^3x\,\overline\Psi\gamma^3\gamma_5
 D^{\mu_1}\ldots D^{\mu_{n-1}} \Psi,
\label{spinop}
\end{eqnarray}
which are the familiar operators of the operator
product expansion for deep inelastic scattering
(see for example \cite{Ahmed}). 
\\
In view of
our aim to calculate the
t evolution of the functions $L_q(t)$ and $L_g(t)
$ {\em generalized to higher moments},
$L_q^n(t)$ and $L_g^n(t)$, we expand by
analogy the quark and
gluon orbital angular momentum operators:
\begin{eqnarray}
 &\widehat {L}_q &= \int d^3x\, i\overline\Psi
 \gamma^+(x^1\partial^2-x^2
 \partial^1)\Psi \nonumber\\
 \longrightarrow \ &\widehat {L}_q^{\mu_1\ldots
  \mu_{n-1}} &= \int d^3x\, i
 \overline\Psi \gamma^+(x^1\partial^2-x^2
 \partial^1) D^{\mu_1}\ldots D^{
  \mu_{n-1}}\Psi 
\label{Lqerw} \\ 
 &\widehat {L}_g &= \int d^3x\, \partial^+A^{j,a}
 (x^1\partial^2-x^2 \partial^1) A^{j,a}
 \nonumber\\ \longrightarrow \ &\widehat {L}_g^{
  \mu_1\ldots\mu_{n-1}} 
 &= \int d^3x\, \partial^+
 A^{j,a_1} (x^1\partial^2-x^2\partial^1)
 D^{\mu_1,a_1a_2}
 \ldots D^{
  \mu_{n-1},a_{n-1}a_n} A^{j,a_n}
\nonumber\\
\label{Lgerw}
\end{eqnarray}
$D^\mu$ is the covariant derivative
$\partial^\mu-igT^aA^{\mu,a}$ (respectively
$D^{\mu,ab}=\delta^{ab}\partial^\mu-gf^{abc}A^{
 \mu,c}
$).
\section{Calculation of the evolution coefficients}  
Due to operator mixing, we have to calculate
the $4\times 4$ anomalous dimension matrix for the
set of operators $\widehat{\vec J}=(
\widehat{S}^{\mu_1\ldots \mu_{n-1}}_q,
\widehat{S}^{\mu_1\ldots\mu_{n-1}}_g,\widehat
{L}_q^{
 \mu_1\ldots
  \mu_{n-1}},\widehat {L}_g^{\mu_1\ldots
  \mu_{n-1}})$, which is given by

\begin{equation}
 \gamma_{ij}=\left(Z^{-1}\frac {\partial Z}{\partial
  \ln(\Lambda^2)}\right)_{ij}=-\frac {\partial
  (Z^{-1}_{O(g^2)})_{
   \ ij}}{\partial\ln
  \Lambda^2} + O(g^4),
\label{anomaledim}
\end{equation}
where the matrix $Z$ denotes the operator
renormalization matrix, defined through
\begin{equation}
 \widehat {\vec J}^{\ \mbox{ren}}_i=(Z^{-1})_{ij}
  \widehat {\vec J}^{\ \mbox{unren}}_j.
\label{Z}
\end{equation} 
To determine $Z^{-1}$ we
inspect the insertion of our generalized orbital
angular momentum operators (\ref{Lqerw},
\ref{Lgerw}) into off-forward matrix elements 
\begin{equation}
 \langle p',s|\widehat {L}_{q,g}^{\mu_1\ldots
  \mu_{n-1}}|p,s\rangle
\label{offfor}
\end{equation}
($|p,s\rangle$ is a quark or gluon helicity state) and
calculate their divergent parts. In the
following, we restrict
ourselves to LO, i.e. only 1-loop-graphs are
considered, and work entirely in LCC and the
LC-gauge \cite{Kogut}. In an interacting
theory, the momentum and spin eigenstates
can be expanded pertubatively order by order in
the coupling constant $g$
\begin{eqnarray}
 |p,s\rangle &=& |p,s\rangle_0+\frac {1}{p^--H_0
  +i\epsilon} H_{\mbox{int}} |p,s\rangle_0 + O(g^2) \nonumber\\
 &=& |p,s\rangle_0\quad+\quad |p,s\rangle_1 \quad +\quad O(g^2).
\label{entw}
\end{eqnarray}
Here $H_{\mbox{int}}$ is just the quark-gluon or 
three-gluon interaction vertex.
As we are only interested in the twist-2 part of the $Q^2$-evolution 
we have  only to deal with the
\mbox{($\mu_1=+,\ldots,\mu_{n-1}=+$)}
 components of the operators.
Graphs like
figure 1e, which come from the covariant
derivatives in (\ref{Lqerw},\ref{Lgerw}), do not
contribute in the light cone gauge \cite{Kim}. An
explicit calculation
shows that the matrix elements have the following
general structure
(here $ F(p,p';n)$ and $ G(p,p';n)$ just stand 
symbolically for the individual contributions listed in Eq. (26)-(29)) 
\begin{eqnarray}
 &&\left(\langle p',s| \widehat L_{q,g}^{\overbrace {+
   \cdots +}^{n-1}}|p,s\rangle_1 , \Phi(p')\right) = 
 \nonumber\\ 
 &&=\left((2\pi)^3\left[\bigg(ip^2\frac{\partial}{\partial
  p'^1}-ip^1\frac{\partial}{\partial p'^2}\bigg)
 \delta^3(\vec p\,'-\vec p)\right] F(p,p';n),\Phi(p')
 \right)
 \nonumber\\  
 &&=-\left((2\pi)^3
 \delta^3(\vec p\,'-\vec p)\bigg(ip^2\frac{\partial}{\partial
  p'^1}-ip^1\frac{\partial}{\partial p'^2}\bigg) F(p,p';n),\Phi(p')\right)
 \nonumber\\ 
 &&\quad-
 \left((2\pi)^3
 \delta^3(\vec p\,'-\vec p)F
 (p,p';n),\bigg(ip^2\frac{\partial}{\partial
  p'^1}-ip^1\frac{\partial}{\partial p'^2}\bigg)\Phi
 (p')\right).
 \label{general}
 \end{eqnarray}
Observing that
\begin{eqnarray}
 &&\bigg(\langle p',s| \widehat L_{q,g}^{+
   \cdots +}|p,s\rangle_0 , \Phi(p')\bigg)
 =
 \nonumber\\ 
 &&= -\left(
 (-ip^+)^{n-1}2p^+ (2\pi)^3
 \delta^3(\vec p\,'-\vec p),\bigg(ip^2\frac{
  \partial}{\partial
  p'^1}-ip^1\frac{\partial}{\partial p'^2}\bigg)\Phi
 (p')\right) 
 \label{Lnorm}
\end{eqnarray}
and
\begin{eqnarray}
 &&\left((2\pi)^3\delta^3(\vec p\,'-\vec p)\bigg(ip^2\frac{
  \partial}{\partial
  p'^1}-ip^1\frac{\partial}{\partial p'^2}\bigg)F(p,p';n),
  \Phi(p')\right) = 
  \nonumber\\ 
  && =
 \bigg(G
 (p,p';n)
 \langle p',s|\widehat
 S^{+\cdots +}_{q,g}|p,s\rangle_0,\Phi(p')\bigg)
\label{Serwartung}
\end{eqnarray}
the elements of $Z^{-1}$ are easy to determine:
The first term on the r.h.s. of Eq. (\ref{general}) gives the
contribution from the
spin operators and the second the contribution from
the orbital angular momentum operators to the
renormalization of $\widehat L_{q,g}^{+\cdots +}$.
\\
With the definitions 
\begin{eqnarray}
 \label{defop0}
 \langle\widehat{Op}\rangle_0^{q,g} &=& \left(\langle p',s|
 \widehat{Op}|p,s
 \rangle_0^{quark,gluon} , \Phi(p')\right) \\ \langle\widehat{Op}
 \rangle_1^{q,g} &=& \left(
 \langle p',s| \widehat{Op}|p,s
 \rangle_1^{quark,gluon} , \Phi(p')\right)
 \label{defop1}
\end{eqnarray}
the divergent parts of the matrixelements are (figures 1a-1d)
\begin{eqnarray}
  \langle\widehat L_{q}^{+\cdots +}\rangle_1^q &=& \frac {
  \alpha}{2\pi}\ln(
  \frac{\Lambda^2}{\mu^2}) \bigg\{ \bigg[  C_F
  \int^1_0dx\, x^{n-1}(\frac{x
   +x^3}{1-x})-\underbrace{C_F\int^1_0dx\,(\frac{1
    +x^2}{1-x})}_a
   \bigg] \langle\widehat L_{q}^{+
   \cdots +}\rangle_0^q \nonumber\\ &&+\qquad C_F
  \int^1_0dx\,
   x^{n-1}(x^2-1)\langle\widehat S_{q}^{+
   \cdots +}\rangle_0^q \bigg\},    
\label{Lqqdiv} \\ \langle\widehat L_{q}^{+\cdots +}
\rangle_1^g &=& \frac {\alpha}{2\pi}\ln(
  \frac{\Lambda^2}{\mu^2}) \bigg\{ n_f
  \int^1_0dx\, x^{n-1}(x^3+(1-x)^2x) \langle\widehat L_{g}^{+
   \cdots +}\rangle_0^g \nonumber\\ &&+\qquad n_f
  \int^1_0dx\,
   x^{n-1}(-2x^3+4x^2-3x+1)\langle\widehat S_{g}^{+
   \cdots +}\rangle_0^g \bigg\},
\label{Lqgdiv} \\ \langle\widehat L_{g}^{+\cdots +}
\rangle_1^q &=& \frac {\alpha}{2\pi}\ln(
  \frac{\Lambda^2}{\mu^2}) \bigg\{ C_F
  \int^1_0dx\, x^{n-1}(1+(1-x)^2) \langle\widehat L_{q}^{+
   \cdots +}\rangle_0^q + \nonumber\\ &&+ \qquad
  C_F \int^1_0dx\,
   x^{n-1}(-x^2+3x-2)\langle\widehat S_{q}^{+
   \cdots +}\rangle_0^q \bigg\},   
\label{Lgqdiv} \\ \langle\widehat L_{g}^{+
 \cdots +}
\rangle_1^g &=& \frac {\alpha}{2\pi}\ln(
  \frac{\Lambda^2}{\mu^2}) \bigg\{ \bigg[ 2C_A
  \int^1_0dx\, x^{n-1}\frac{x^4-2x^3+3x^2-2x
   +1}{1-x} \nonumber\\ &&- \qquad \underbrace{\frac{n_f}{2}
  \int^1_0 dx\, (2x^2-2x+1)}_b
  \nonumber\\ &&- \qquad\underbrace{\frac{C_A}{2}
   \int^1_0 dx\, (\frac{x^3
   +3x}{1-x}-(x^2-x-1))}_c \bigg] \langle\widehat L_{g}^{+
   \cdots +}\rangle_0^g \nonumber\\ &&+ \qquad
  2C_A \int^1_0dx\,
   x^{n-1}(x-1)(x^2-x+2)\langle\widehat S_{g}^{+
   \cdots +}\rangle_0^g \bigg\}.
\label{Lggdiv}
\end{eqnarray}
$\Lambda$ is the
UV-cutoff from the transverse momentum
loop integral ($\mu$ is the infrared cutoff). The
terms $a$, $b$ and $c$ correspond to the quark
respectively gluon self energy, coming from
the corresponding field renormalization constants
$Z^{-1}_{q,g}$ which are implicitely
included in
the matrix elements. In
combination with (\ref{anomaledim}) and the definition
$\frac{\alpha(t)}{2\pi}A_{ij}=\gamma_{ij}$, we obtain
the generalized and completed evolution
equation for the n-th moments of spin and orbital
angular momentum distributions
\begin{eqnarray}
&&\frac{d}{dt}
\left(
\begin{array}{llll}
 \Delta\Sigma^n \\
 \Delta g^n \\
 L_q^n \\
 L_g^n \\
\end{array}
\right)
= \frac{\alpha(t)}{2\pi} 
\left(
\begin{array}{cc}
 A^n_{SS} & A^n_{SL} \\
 A^n_{LS} & A^n_{LL}
\end{array}
\right)
\left(
\begin{array}{llll}
 \Delta\Sigma^n \\
 \Delta g^n \\
 L_q^n \\
 L_g^n \\
\end{array}
\right)
\\ 
\nonumber\\
&&A^n_{SS}=\left(
\begin{array}{cc}
 C_F\left[\frac{3}{2}+\frac{1}{n(n+1)}-2\sum^n_{j
 =1}\frac{1}{j}\right] & n_f\left[\frac{n-1}{n(n
 +1)}\right] \\
 C_F\left[\frac{n+2}{n(n+1)}
\right] & 2C_A\left[\frac{11}{12}-
\frac{n_f}{6C_A}+\frac{2}{n(n+1)}-\sum^n_{j=1}\frac{1}{j}\right]
\end{array}
\right)
\nonumber \\
&&A^n_{SL}=\left(
\begin{array}{cc}
 0 & 0 \\
 0 & 0
\end{array}
\right)
\nonumber \\
&&A^n_{LS}=\left(
\begin{array}{cc}
 -2C_F\left[\frac{1}{n(n+2)}\right] & n_f\left[
\frac{n^2+n+6}{n(n+1)(n+2)(n+3)}\right] \\
 -C_F
\left[\frac{n+4}{n(n+1)(n+2)}\right] & -4C_A
\left[\frac{n^2+4n+6}{n(n+1)(n+2)(n+3)}
\right]
\end{array}
\right)
\nonumber \\
&&A^n_{LL}=\left(
\begin{array}{cc}
 C_F\left[\frac{3}{2}-\frac{2n
  +3}{(n+1)(n+2)}-2
\sum^n_{j=1}\frac{1}{j}\right] & n_f\left[
\frac{n^2+3n+4}{(n+1)(n+2)(n+3)}\right] \\
 C_F
\left[\frac{n^2+3n+4}{n(n+1)(n+2)}\right] & 2C_A
\left[-\frac{n^3+3n^2-6}{n(n+1)(n+2)(n+3)}-
\frac{n_f}{6C_A}+\frac{11}{12}-\sum^n_{j=1}
\frac{1}{j}\right]
\end{array}
\right)
\nonumber
\end{eqnarray}
$A^n_{SS}$ is the known DGLAP-matrix of the spin
dependent case. To our knowledge the  coefficient matrices
$A^n_{LS}$ and $A^n_{LL}$ were not calculated before
except for $n=1$. 
$A^n_{LL}$ turns out to be equal
to the spin-independent DGLAP-matrix with $n
\rightarrow n+1$ (\cite{AP},\cite{A}).
\\
\\
The results for $A^n_{SL}$ and $A^n_{LL}$ can be interpreted in the
following manner. Angular momentum can only be sensibly defined
between at least two particles. On the level of the twist-2
DGLAP equations which describe the splitting of one parton into two 
the angular momentum of the initial parton cannot play any special
role. It is only relevant for the final state, which contains
two partons. Here it  allows for a spin flip. 
The momentum (longitudinal and transverse part) of the incoming
parton is just split between  
the daughter partons with the ratio $x/(1-x)$. Therefore 
$A^n_{LL}$ turns out to be the usual spin-independent 
results just shifted by 1 in $n$, because the angular momentum 
operator contains the momentum as a factor.\\
Following the same arguments one has to conclude that 
the angular momentum carried by the initial parton
(relative to the nucleon center of momentum) does
not couple to the final state spins. Therefore 
$A^n_{SL}$  has to be zero.\\ 
\\
To check our results we repeated our calculation following 
the manner in which 
the splitting functions $P(x)$ are usually derived and
just inserting the angular momentum operators 
Eq. (\ref{deltasigma})-(\ref{Lg}).
In this case some care is needed when the
$\frac{1}{1-x}$ singularities of the diagonal
splitting functions are replaced by the 
$(\frac{1}{1-x})_+$ distributions \cite{AP}.
%
Let us note that Eq. (\ref{summezerlegt})  implies 
\begin{equation}
 \frac{d}{dt}\frac {\bigg(\Delta\Sigma(t)/2 + \Delta g(t) + L_q(t) + L_g(t),
  \Phi(P')\bigg)}{2P^+(2\pi)^3\Phi(P)} = 0
\label{nebenbed}
\end{equation} 
which is just another check.
\\
\\
The estimation of the phenomenological implications
of our results requires numerical simulations with
a great variety of initial angular momentum distributions.
These studies will necessarily invoke a substantial amount of
model-building, while the results presented in this contribution 
are valid in general.\\
\\
{\bf Acknowledgments}
\\ 
We are very grateful to Xiangdong Ji for many helpful 
discussions and for making his notes for Ref. \cite{JiI} available to us.


\clearpage

\begin{figure}[t]
\epsfig{file=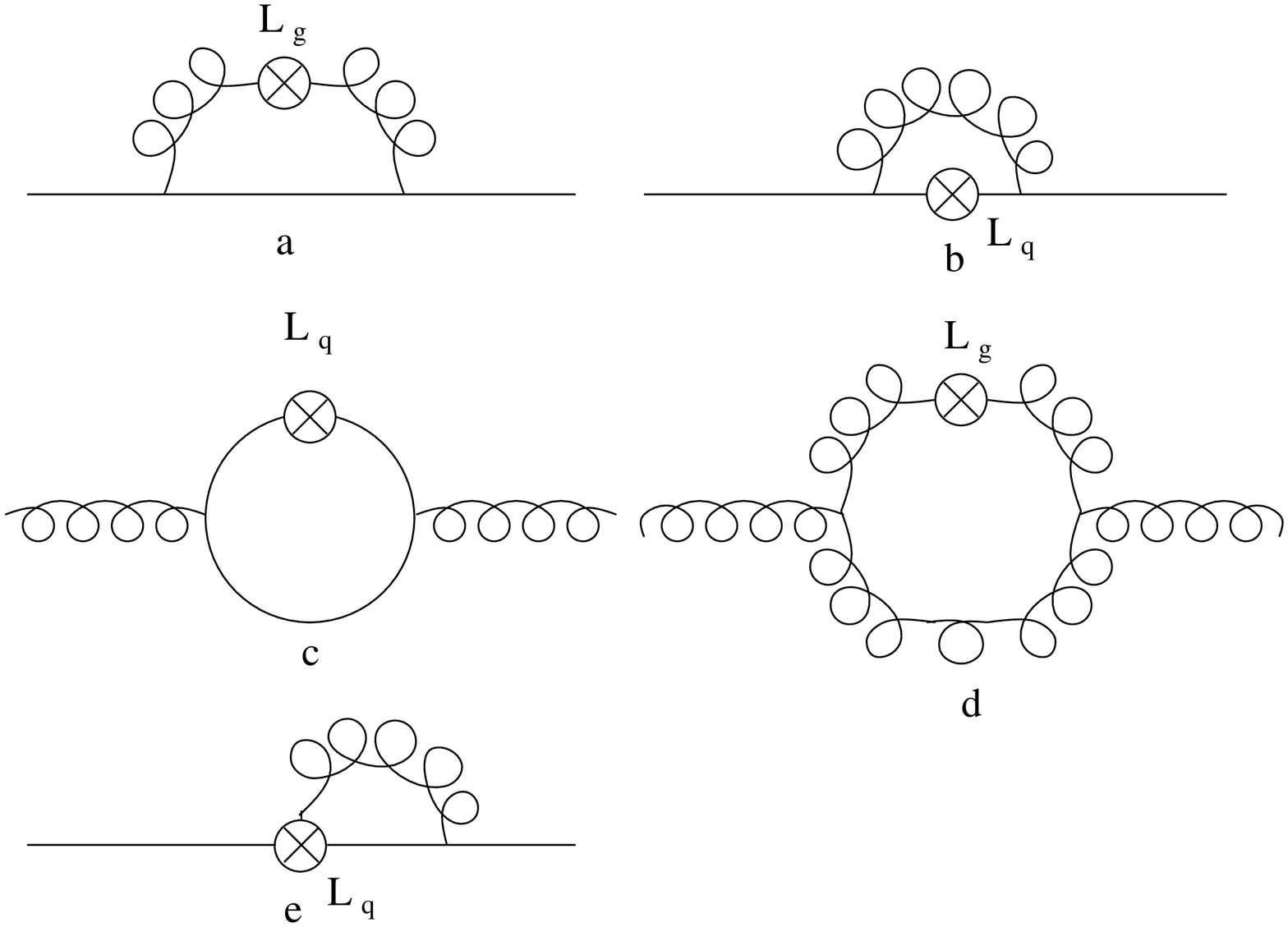, height=10cm}
\caption{loop-graphs with inserted operators}
\end{figure}

\end{document}